\documentclass[a4paper,10pt]{revtex4}

\usepackage{amsmath,amssymb,amsfonts}
\usepackage{algorithmic}
\usepackage{graphicx}
\usepackage{textcomp}
\usepackage{xcolor}

\renewcommand{\theequation}{\thesection.\theequation}
\numberwithin{equation}{section}

\begin{document}

\title{ 
High sensitivity electrochemical DNA sensors for detection of somatic mutations in 
FFPE 
samples
}

\author{Valentina  Egorova}
\email{egorova_vp@bspu.by}
\affiliation{Belarusian State Pedagogical University, Minsk, Belarus  } 


\author{Halina Grushevskaya}
\email{grushevskaja@bsu.by}
\affiliation{Physics Faculty, Belarusian State University, 4 Nezavisimosti Ave., 220030 Minsk, Belarus}

\author{Nina Krylova}
\email{nina-kr@tut.by}
\affiliation{Belarusian State  Agrarian Technical University, Minsk, Belarus}

\author{Igor Lipnevich}
\email{lipnevich@bsu.by}
\affiliation{Belarusian State University, Minsk, Belarus }


\author{Egor Vaskovtsev}
\email{vaskovcev@mail.ru} \affiliation{Belarusian State Pedagogical University, Minsk, The Republic of Belarus }


\author{Andrei Babenka}
\email{labmdbt@gmail.com}
\affiliation{Belarusian State Medical University, Minsk, Belarus }


\author{Ina Anufreyonak}
\email{klaksa_06@mail.ru}
\affiliation{N. N. Alexandrov National Cancer Center of Belarus, Lesnoy, Belarus}

\author{Siarhei Smirnov }
\email{rustledeath24@gmail.com}
\affiliation{N. N. Alexandrov National Cancer Center of Belarus, Lesnoy, Belarus}

\begin{abstract}
We offer new high-performance label-free electrochemical
impedimetric DNA sensors of non-faradaic type. The DNA sensors
based on a platform of crystalline carbon nanotube (CNT) arrays
are fabricated by the Langmuir--Blodgett (LB) deposition
technique. The CNT  arrays are suspended on a nanoporous anodic
alumina (aluminium oxide, AOA) support.
Single-stranded (ss) 19- and 20-base oligonucleotides and double-stranded (ds) nucleotide
sequences were used as probes for chosen molecular target -- human
KRAS (Kirsten Rat Sarcoma viral oncogene homolog) gene. Genomic deoxyribonucleic acids
(DNAs) were isolated from placenta of healthy donors and FFPE
(formalin-fixed, paraffinembedded) samples of tumor tissue. Raman
spectral analysis and electrochemical dielectric spectroscopy were
used to sequence the target DNAs. The optical and electrochemical
detection (variants of DNA sequencing) were based on  a screening
effect that grows after the DNA homoduplex formation. We
demonstrated that such technologies are very sensitive and allow
to detect attomolar DNA concentrations and less. As result, the
KRAS exon 2, codon 12, c.35G$>$A mutation was successfully
discriminated in human genomic DNA isolated  from FFPE colorectal
cancer tumor tissue samples.
\end{abstract}

\maketitle

{\bf Keywords:}
DNA-carbon nanotube hybrid, conducting polymer, electric-field
screening effect, temperature effects, DNA melting, attomolar DNA sensor, diagnostics







\section{Introduction}
The modern approaches to tumor therapy cannot exist without
high-quality molecular diagnostics. Currently, the  U.S. Food and
Drug Administration (FDA) has approved dozens of molecular tests
aimed at choosing targeted therapy for tumors of various
nosological forms: colorectal cancer, breast cancer, lung cancer,
etc. \cite{FDA,Nakajima2022,Arora2021}
Most often, FFPE sections are used for
solid tumors. This allows the operator to select the area of
interest for analysis as accurately as possible
\cite{Panchal2020}. The quality of the extracted DNA varies
greatly from sample to sample. This has a strong influence on the
results of molecular tests \cite{Kapp2015}
.
During the preparation of FFPE blocks, the DNA undergoes
fragmentation. As a result, the number of sites for annealing of
oligonucleotide primers is reduced. This is a key parameter for
the implementation of PCR-based techniques. When extracting and
purifying DNA from FFPE sections, the degree of fragmentation also
increases, further reducing sample quality. In addition, the
effectiveness of PCR-based methods is also affected by the
presence of inhibitors that enter the DNA solution in small
concentrations. This may not stop the reaction, but reduce its
effectiveness. When there are very few target sequences in
solution, this can be crucial and lead to diagnostic errors \cite{Lu2018}
.
One possible solution to the problem is the use of DNA sensors and 
 electrochemical impedance spectroscopy (EIS). Since this is not a
standard approach, it has many directions and mechanisms \cite{Li2021}
.
DNA sensors have at least two advantages. First, they are much
less dependent on the presence of inhibitors in the DNA sample
compared to methods using enzymes. Second, a shorter DNA
fragment/probe is used to recognize the target sequence. This
increases the number of fragmented DNA sequences suitable for
recognition \cite{Tadimety2022}
.
Currently, there are no ready-made solutions that would fully meet
the needs of the clinical diagnostics market. Many DNA sensor
technologies have a number of disadvantages so far. Their
improvement and optimization for work in real clinical conditions
is an important and priority area of research
\cite{Nesvet2021,Kim2021}
.
To date, the development of low-temperature genotyping also is in high demand \cite{LiEtAl}.

Graphene materials and, in particularly,  rolled atomically-thin carbon layers called carbon nanotubes
are promising materials for development of high performance electrochemical and optical
transducers of DNA-hybridization signals. The transducers
operating on plasmon resonance effects in the nanomaterials are
capable to perform the nucleotide sequencing of samples with
ultralow DNA-target concentrations: attomoles and less
\cite{Quinchia2020}.
Massless graphene charge carriers are high mobile even in
comparison with electrons of the ordinary metals.
Physics of the graphene charge carriers residing in valleys $K,K'$ of graphene Brillouin zone
 is  Dirac physics. Chirality of the massless pseudo Dirac fermions prevents annihilation of electron-hole pairs.
But, the charge transport in graphene also is featured by signs of a non-Abelian statistics
of the Majorana quasiparticles with a nontrivial topology
\cite{My2022QuantumReport,myNPCS18-2015,NPCS18-2015GrushevskayaKrylovGaisyonokSerow,
Taylor2016,Grush-KrylSymmetry2016,our-symmetry2020}.

However, the electrical and optical properties of nanostructured graphene-like materials strongly depend
on accidental environmental impurities  that leads to
wide spread of characteristics 
of the graphene-based devices. Moreover, when scattering and
obliquely tunneling through  the environmental  defects the
graphene charge carriers  can be confined 
as Klein resonances \cite{My2021PhysRevB}.
 The electrostatical confinement is  predicted within the  topologically-nontrivial graphene model
\cite{My2022JNPCSKleinRes}.
The  fine tuning of the Klein resonances is a challenge. To design a graphene-based material
with specified characteristics 
one needs the Langmuir--Blodgett  technique which allows to
fabricate highly ordering defect-free layered nanocomposites which
include few-walled carbon nanotubes (FWCNTs).

Immobilization of DNA on the graphene surface or using of high-conductive oligomer improve electrical
transport along DNA (see \cite{XuEtAl2006,Abdullin2009,Benvidi2016,Rasheed2017} and
references therein). But, the specificity of the probe DNA is
attenuated at its immobilization.   Therefore, the construction of
impurity-insensitive label-free PCR-free transducers based on
graphene materials for detection of single-molecule hybridization
signal is complicated and unsolved problem.

The homoduplex formation can be indicated through emerging
transition from the fully ordered double-helix state of ssDNA
probe--ssDNA target duplexes  to the metastable state of the
complexes at the temperature increase. Although the extended
linear conformation of the  probe and target ssDNA molecules
entering the homoduplex  remains  in the metastable state, the molecules
separate from each other in a diffusion process because they are not linked by hydrogen bonds and, correspondingly,
almost do not interact with each other. As a result, a helix-coil transition
happens at the temperature alternation. The melting of long dsDNA at
temperature increase is a phase transition of the first order
\cite{Yoshikawa1996}.
 The temperature of the DNA transition (so-called melting temperature $T_m$) from the helicoidal form to the linear
one  depends on the length of the ssDNA molecules. The phase transitions at DNA melting are detected  in a wide temperature
region because $T_m$ depends significantly on the oligonucleotide
length and base composition. Developing DNA sequencing methods one needs to narrow  the temperature region.
The value of $T_m$ is also necessary to design the DNA probes as its experimental
determination  reduces the time needed for construction of DNA probes with required parameters.

When creating a network the complexification between
highly-ordered carbon-nanotube arrays and dsDNA target molecules
  do not impair but improve the ability of the FWCNT arrays to
screen electric fields \cite{Babenko2020,Grushevskaya2018S}. The
high-conductive oligomers with conjugated double bonds and FWCNTs
form a network also, enhancing the efficiency of the screening by
the highly-ordered FWCNT arrays
\cite{Grushevskaya2019,Egorova2020}.

In the paper we intend to discriminate a mutation status of KRAS
gene. When activated, a KRAS protein is involved in the
dephosphorylation of guanosine triphosphate (GTP) to guanosine
diphosphate (GDP), after which KRAS is turned off. KRAS-gene
single base mutations called  single-oligonucleotide polymorphysms
(SNPs) most often occur in exon 2, codon 12 of the KRAS gene with
the nucleotide change c.35G$>$A (substitution, position 35, G$\to$A)
resulting in the amino acid change   p.G12D  (substitution --
missense, position 12, Gly12Asp (G$\to$D)) We offer to detect the
genotype on a change of the degree of shielding by a sensor
electrode coating in a result of the DNA homoduplex formation on
the FWCNT which surface was decorated  by conducting oligomers of
thiophene-pyrrole series entering cyclic organometallic complexes.
The screening efficiency for the network created by linkage of
hybrids between the
homoduplexes and decorated CNTs   
grows but a transfer of electrons along the organometallic complexes
is broken 
because when penetrating through nanocavities of the nanocyclic compound the target DNA
disturbs perfect location of S atoms in the thiophene groups  relative to
the nitrogen atoms of the pyrrole groups 
 \cite{Babenko2020,Egorova2020}.
We shall elucidate the origin of light-scattering enhancement for dsDNA deposited on the
graphene plane   as a result of intensive graphene-charge-density
oscillations in resonance with vibrations of the dsDNA molecular
groups, which are distant enough from the graphene surface.
We shall experimentally show that the light scattering in the dsDNA
homoduplexes deposited on rolled graphene monolayers is enhanced
by graphene charge carriers of  the pseudo-Majorana type rather than the pseudo-Dirac type ones.

A goal of the paper is to develop a high-performance
low-temperature PCR-free electrochemical sensor for KRAS
c.35G$>$A, p.G12D genotyping in FFPE tumor samples. The sensor
operates on
 screening  near-electrode double layer
 by crystalline FWCNT arrays. We shall show that depositing oligonucleotides of several types on the electrode surface
 one gets an EIS DNA chip for discriminating between two alleles of KRAS-oncogene.
 The KRAS-oncogene sequencing will be verified by a DNA-melting control.

\section{Materials and methods}
\subsection{Reagents}


Genotyping was performed in placental DNA, nuclear DNA of C6-line
rat glioma cells, and tumor DNA denoted by dsDNA$_{\mathrm{pl}}$, dsDNA$_{\mathrm{C6}}$, and
dsDNA$_{\mathrm{CRC}}$, respectively.
The following SNP: c.35G$>$A site, p.G12D,  located in the
second exon of the human KRAS gene (NCBI/Gene KRAS genomic DNA sequence NC\underline{ }000012.12;
NCBI/Gene KRAS mRNA var d sequence NM\underline{ }001369787.1), was used as a molecular target.
Clinical samples were collected at the N.~N.~Alexandrov
National Cancer Center of Belarus.  The study included 10~FFPE
samples obtained from patients with advanced colon cancer. For
isolation, 2--3 sections of the paraffin block were used.
Deparaffinization was performed with xylene followed by washing
with ethanol (95\%). Total DNA isolation was performed using the
QIAamp DNA FFPE Tissue Kit (Qiagen) according to the
manufacturer's protocol.



The C6-line rat glioma cells were cultured in standard conditions. The  native
dsDNA$_{\mathrm{pl}}$ was isolated from placenta tissue of healthy donors.
RNA and proteins contents in the high purity dsDNA$_{\mathrm{pl}}$ and dsDNA$_{\mathrm{C6}}$ (1.03~mg/ml
in $10^{-5}$~M Na$_2$CO$_3$ buffer medium) 
were less that 0.1~\%~(optical density ratio $D_{260}/D_{230} =
2.378$ and $D_{260}/D_{280} = 1.866$, respectively).

The single stranded  19- and 20-base oligonucleotides, short primers KRAS$_w$ and KRAS$_m$, and
two double-stranded (toehold exchange) probes, P3/W3 and ssDNA$_{lp}$/ssDNA$_{Wm}$, were utilized as DNA probes
(see Table~\ref{tab1}).  The perfect-matched (KRAS$w$) and single-mismatched (KRAS$m$) probes are
complementary to  the KRAS-gene  wild-type (non-mutant) and mutant-type  nucleotide sequences, respectively. The
 P3/W3 toehold dublex consisted of the two 35-base and 28-base oligonucleotides.
The longer capture oligonucleotide, W3, is complementary to the wild KRAS gene. The four inosines
replace guanines in the shorter chain, P3, being a protector of the
capture oligonucleotide ``W3''.
The ssDNA$_{lp}$/ssDNA$_{Wm}$ toehold dublex consisted of the two 47-base and 40-base oligonucleotides.
The longer capture oligonucleotide, ssDNA$_{Wm}$, is complementary to the KRAS gene with the SNP.
The five inosines replace guanines in the shorter chain, ssDNA$_{lp}$, being a protector of the
capture oligonucleotide, ssDNA$_{Wm}$.

\begin{table}[htbp]
\caption{Sequences of all the oligonucleotides used in this
study.}
\begin{center}
\begin{tabular}{|p{40pt}|p{340pt}
|
}
\hline
Name & Sequence structure ($5' \to 3' $) 
 \\
\hline N$_3$ & TGGTGGCGTAGGCAAGAGTGCCTTGACGATACAGC  
\\
\hline P$_3$ &  GCTGTATCGTCAAGGCACTCTTGCCTACGCCACCA 
\\
\hline  W$_3$ &  TGGTGICGTAGICAAGAITGCCTTIACG 
\\
\hline  M$_3$ &  TG{\bf A}TGGCGTAGGCAAGAGTGCCTTGACGATACAGC  
\\
\hline KRAS$_w$ & GTTGGAGCTGGTGGCGTAG  
\\
\hline KRAS$_m$ & AGTTGGAGCTG{\bf A}TGGCGTAG 
\\
\hline 
ssDNA$_{lp}$ &
AATAAGGAGGCACTCTTGCCTACGCCATCAGCTCCAACTACCACAAG
\\
\hline 
ssDNA$_{Wm}$ &
CTTGTGITAGTTGGAGCTGATGICGTAGICAAIAGTICCT 
\\
\hline
ssDNA$_{lm}$ &
CTTGTGGTAGTTGGAGCTG\textbf{A}TGGCGTAGGCAAGAGTGCCTCCTTATT 
\\\hline
polyG & GGGGGGGGGGGGGGGGGGGGGGGGG 
\\
\hline
\end{tabular}
\label{tab1}
\end{center}
\end{table}

The  toehold exchange probes were fabricated by annealing the
protector  P3 or ssDNA$_{lp}$ strands mixed  in the ratio 2:1
with the complementary W3 or ssDNA$_{Wm}$ strands, respectively,
in TE buffer at $37 \pm 0.1~^\circ$C. The 35-base oligonucleotides
denoted by ``N3'' and ``M3'' in Table~\ref{tab1} are the
perfect-matched and single-base mismatched KRAS-gene sequences and
have been used as  wild- and mutant-type  model target ssDNAs,
respectively. The 47-base oligonucleotide denoted by
``DNA$_{lm}$'' in Table~\ref{tab1} is the single-base mismatched
KRAS-gene sequence and has been used as mutant-type  model target
ssDNA.

A DNA chip was  fabricated with the two oligonucleotides: KRAS$_m$
and P3/W3, or P3/W3 and 25-base guanine oligonucleotide (polyG),
or KRAS$_w$ and ssDNA$_{lp}$/ssDNA$_{Wm}$. The hybridization
reaction lasted 30 minutes. All DNA probes were purchased from
``Primetech ALC'' (Minsk, Belarus).

The carboxylated and stearic-acid-functionalized FWCNTs under
2.5~nm in diameter have been decorated by the cyclic complexes
(Fe(II)DTP) of Ce and/or high-spin octahedral Fe(II) with
3-hexadecyl-2,5-di(2-thienyl)-1H-pyrrole (H-DTP,
H-dithienylpyrrole) ligands 
\cite{my2013JModPhys}. The H-DTP is an amphiphilic conducting oligomer of
thiophene-pyrrole derivatives. An alkyl 16-link hydrocarbon chain
$R=$C$_{16}$H$_{33}$ was chemically bounded to the oligomer.
Inverse micelles of stearic acid with FWCNTs, or DNA--FWCNT, or DNA inside (called micellar FWCNTs,
micellar DNA--FWCNT hybrids, and  micellar DNA) are obtained
by mixing stearic acid and the appropriate reagent dissolved in hexane by the ultrasound treatment.

Salts Fe(NO$_3)_3\cdot 9$H$_2$O, Ce$_2$(SO$_4)_3$ (Sigma, USA),
hydrochloric acid, deionized water with resistivity of
18.2~$M\Omega \cdot$cm were used to preparate subphases.
 All chemical reagents of analytical grade were used
without further purification.

\subsection{Methods}

\subsubsection{ Fabrication of dithienylpyrrole-coated FWCNT arrays}

By employing the LB technique monomolecular crystalline
DNA--FWCNT hybrid layers (DNA-FWCNT hybrid crystalline monolayers) were fabricated by compressing
two-dimensional gas of inverse stearic-acid micelles with the
carboxylated FWCNTs and DNA probe molecules inside. The  hexane solution of the micellar DNA--FWCNT hybrids
 was prelimenary
dropped on deionized-water surface. Then the two DNA--FWCNT hybrid LB-monolayers were formed
and  deposited onto surface of interdigital electrode
structures which were isolated by nanoporous AOA, 10-nm pore
diameter.

Preliminary, five iron-containing  LB-monolayers consisting of the
cyclic complex  (Fe(II)DTP) of octahedral high spin Fe(II) with
DTP ligands were formed. The FWCNTs were decorated by the
nanocyclic organometallic LB-complexes. The FWCNT LB-arrays
decorated by the dithienylpyrrole complexes of octahedral high
spin Fe(II)  host spin-polarized graphene charge carriers
\cite{Grushevskaya2018IJMP,Grushevskaya2016IJMP,Grushevskaya2016}.


\subsubsection{ Electrochemical impedance measurements}

EIS measurements have been made with a BORDO-322 digital oscilloscope (UniTechProm, Minsk, Belarus)
which provides automatic measurements  and mathematical signal processing (fast Fourier transforms) of
harmonic electrical signals.

Interdigital electrode  structures deposited on a glass-ceramic support (pyroceramics) are utilized. $N_e$ pairs,
$N_e=20$ of the aluminium electrodes are arranged in an Archimedes-type spiral configuration.
Every such pair is an ``open type'' capacitor.
To record DNA-hybridization signals a transducer sensitive layer was formed on
the surface of the electrode structures in a following way. The
oligonucleotide sequences--FWCNT hybrid LB-arrays  decorated by the
LB-films of the nanocyclic organometallic complexes were deposited on
the interdigital structure of aluminium electrodes, on the surface
of which the nanoporous anodic alumina layer were previously formed.

To excite harmonic
auto-oscillations of electric current (charge--discharge
(charging) processes in capacitors), the planar capacitive
sensor of interdigital-type
 was connected as the capacitance, $C$, into the relaxation
resistance--capacitance (RC) oscillator (self-excited $RC$-oscillator)
\cite{Abramov2012}.
A self-excitation of an amplifier with a positive feedback occurs in
such RC-oscillator on quasi-resonance frequencies.
  The capacitance $C$ of the sensor entering the measuring RC-oscillating
circuit has been calculated by the formula $C= 1/(2\pi R f)$, where $f$ is a quasi-resonance frequency.

The Bode plots 
were recorded in the deionized water.

\subsubsection{DNA detection}


Double-stranded DNA was denatured at 95~$^\circ$C before detection. The
denatured DNA was hybridized with a DNA probe.

\begin{figure}[htbp]
\centering
(a)\hspace{5cm} (b)\hspace{5cm} (c)\\
\includegraphics*[width=5.5cm, height=6cm, angle=0]{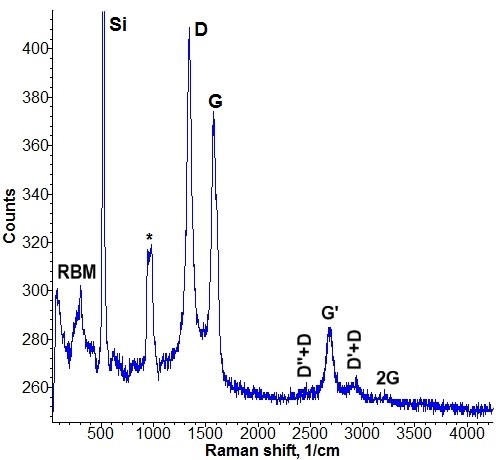}
\includegraphics*[width=5.5cm, height=6cm, angle=0]{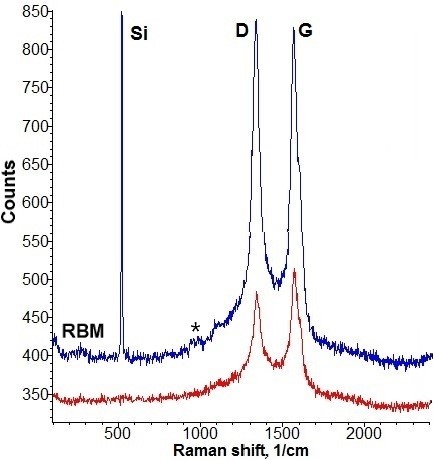}
\includegraphics[width=6.cm,height=6.cm,angle=0]{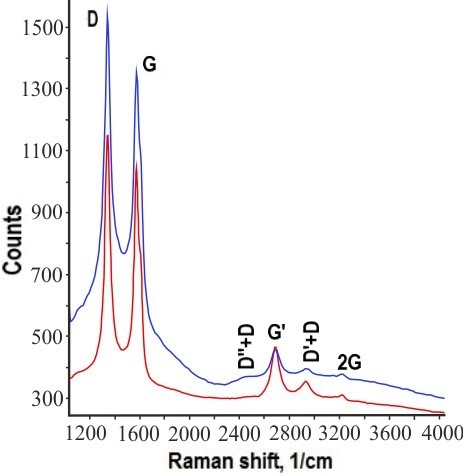}
\caption{ Raman spectra of 
(a) the stearic acid micelles with dsDNA$_{\mathrm{pl}}$--FWCNTs inside, 
(b) the micellar KRAS$_w$--FWCNTs hybrids (red curve in figure b)
and a broken-down (crumpled) LB-monolayer from the stearic acid
micelles with KRAS$_w$--FWCNTs hybrids inside (blue curve in
figure b),
(c) the LB-monolayer formed by the micellar  dsDNA$_{\mathrm{pl}}$-CNT hybrids with (c
,~blue curve) or without (c
,~red curve) propidium iodide. The  micellar KRAS$_w$--FWCNT,
dsDNA$_{\mathrm{pl}}$--FWCNT hybrids, and the KRAS$_w$--FWCNT
hybrid LB-monolayer were deposited on  pure Si supports; the
dsDNA$_{\mathrm{pl}}$-CNT hybrid LB-monolayer was deposited on the
Si support hydrophilized by H-DTP. The Raman spectra were recorded
at laser excitation wavelength of 532~nm; the following laser
powers and collected times were used for the specimen excitation:
20~mW and 1~s (figure~(a)),
3~mW and 10~s (figures~b and c).
 ``*'' denotes a laser mode. }\label{DNACNT}
\end{figure}

\subsubsection{Raman spectroscopy}

Spectral studies in visible range were carried out using a
confocal micro-Raman spectrometer Nanofinder HE (``LOTIS-TII'',
Tokyo, Japan--Belarus) on lasers operating at wavelengths of
473 (DPSS laser), and 532 (DPSS laser)~nm  with power in the range from 0.0001 to 20~mW. The spectra were recorded in
the back-scattering geometry under a $\times$ 50 objective at room (RT) temperature. 
The size for the optical image is
of $7\times 7~\mu$m,  vertical spatial resolution 150~nm, spectral resolution 
 is of up to 0.01~nm.

\begin{figure}[htbp]
\centering
(a) \hspace{5.6cm} (b) \hspace{5.6cm} (c) \\
\includegraphics*[width=5.6cm, height=6cm, angle=0]{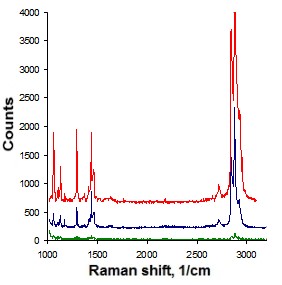}
\includegraphics[width=5.6cm,height=6.cm,angle=0]{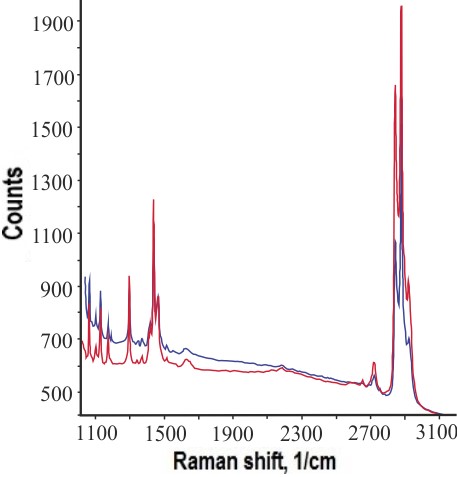}
\includegraphics[width=5.6cm,height=6.cm,angle=0]{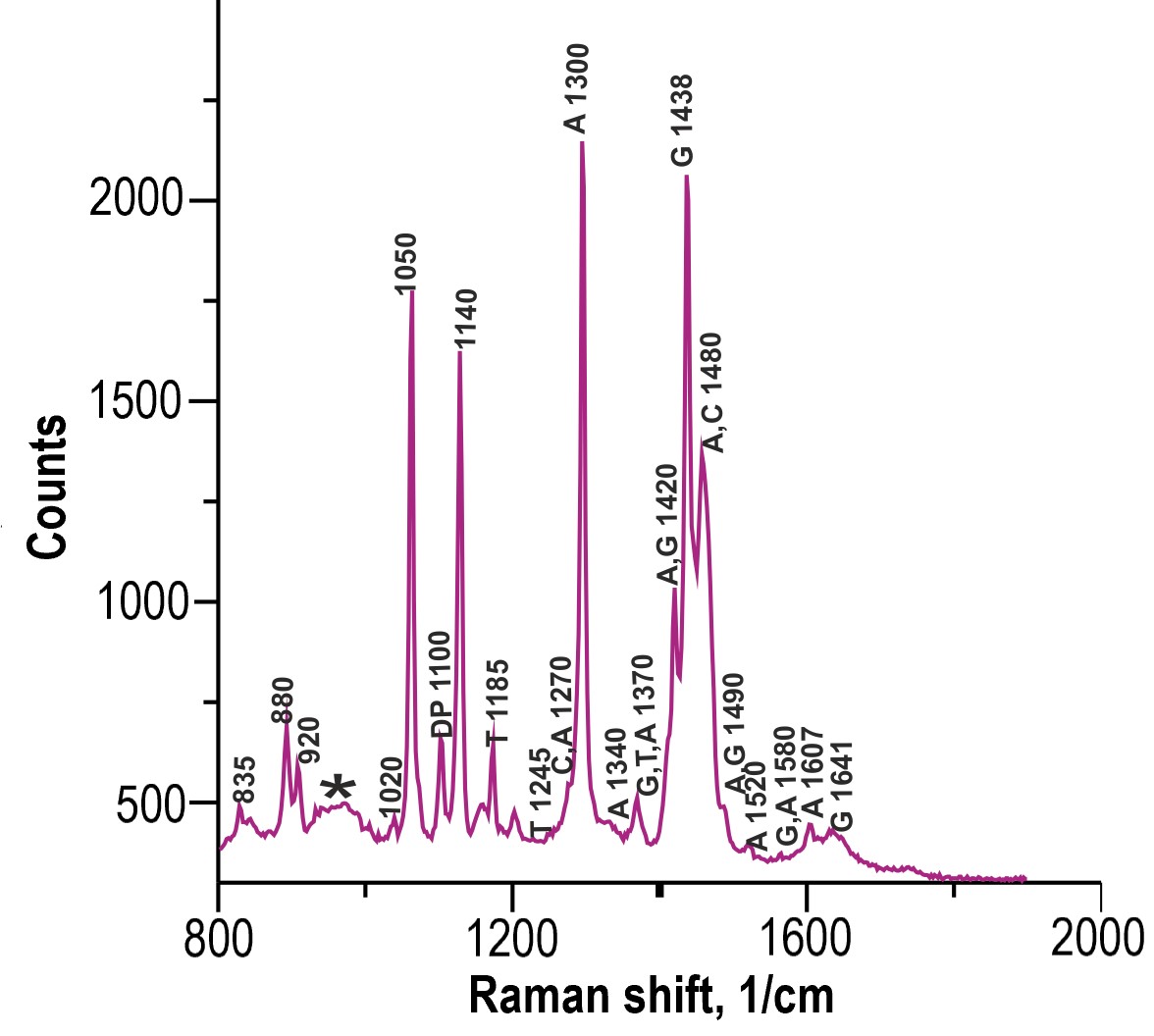}
\\
%
(d) \hspace{7.0cm} (e)\\
\includegraphics[width=6.cm,height=6.cm,angle=0]{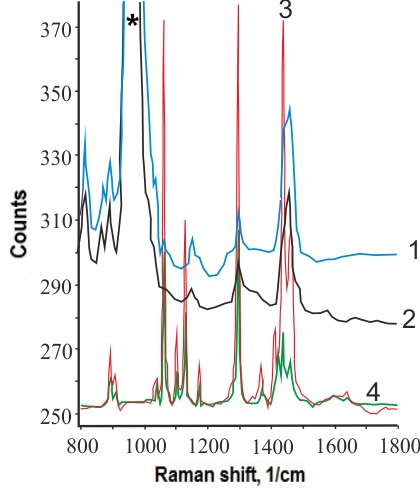}
\includegraphics[width=6.cm,height=6.cm,angle=0]{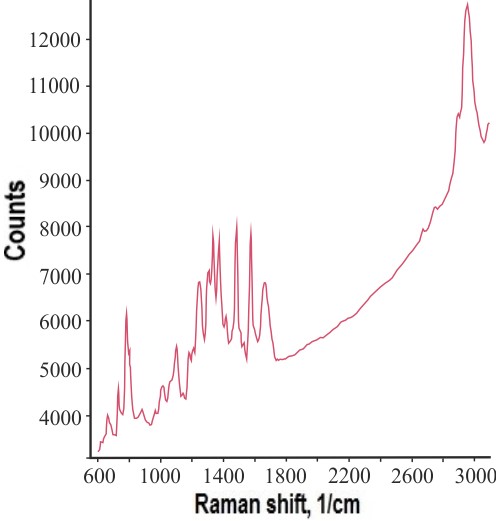}
\caption{ (a) Raman spectra of pure stearic acid inverse micelles
(green curve in figure a), the micelles with dsDNA$_{\mathrm{pl}}$
inside (blue curve in figure a), the micelles with
dsDNA$_{\mathrm{pl}}$--FWCNTs inside (red curve in figure a); the
pure stearic acid micelles and the micelles with
dsDNA$_{\mathrm{pl}}$ and  dsDNA$_{\mathrm{pl}}$--FWCNT hybrids
inside were deposited on pure Si supports; the Raman spectra were
recorded  at laser excitation wavelength of 532~nm; the laser
power and collected time were used for the specimen excitation:
2~mW and 60~s, respectively. (b) Raman spectra of the
micellar nuclear dsDNA$_{\mathrm{C6}}$ (red curve) and the
micellar tumor dsDNA$_{\mathrm{CRC}}$ (blue curve); the
dsDNA$_{\mathrm{C6}}$ and the dsDNA$_{\mathrm{CRC}}$  were
deposited on  pure Si supports; the Raman spectra were recorded at
laser excitation wavelength of 532~nm;
the laser power and collected time were used for the specimen excitation: 2~mW and 60~s. 
(c) A Raman low-frequency part of the spectrum for the inverse
micelles formed in hexane solution of mixture from stearic acid
with dsDNA$_{\mathrm{pl}}$
 and dripped on pure Si. The numbers indicate the characteristic frequencies of dsDNA vibrations;
adenine, guanine, thymine and cytosine
are designated by A, G, T and C, respectively; DP denotes phosphodiester bond.
Laser power, excitation wavelength, and collected time   were of
20 mW,  532~nm, and 3~s, respectively. (d) Raman spectra of   the
micellar dsDNA$_{\mathrm{pl}}$  (blue and black curves 1 and 2,
respectively) and the micellar  dsDNA$_{\mathrm{pl}}$-CNT hybrids
(red and green curves 3 and 4, respectively); the micellar
dsDNA$_{\mathrm{pl}}$  and  dsDNA$_{\mathrm{pl}}$-CNT hybrids were
deposited on  the Si hydrophilized by H-DTP; the Raman spectra
were recorded  at laser excitation wavelengths of 473~nm (curves 1
and 3) and 532~nm (curves 2 and 4), the following laser powers and
collected times were used for the specimen excitation: 8~mW and
1~s (curves 1 and 3), 20~mW and 1~s (curves 2 and 4). (e) A Raman
spectra of  the original  dry placental dsDNA$_{\mathrm{pl}}$
deposited on pure Si support; the spectrum was recorded  at laser
excitation wavelengths of 473~nm; the laser power and collected
time were used for the specimen excitation: 5.76~mW and 60~s.
``*'' denotes a laser mode.} \label{fig1}
\end{figure}

\begin{figure}[htbp]
\centering
(a)   \hspace{6.3cm} (b)\\
\includegraphics[width=6.cm,height=9.cm,angle=0]{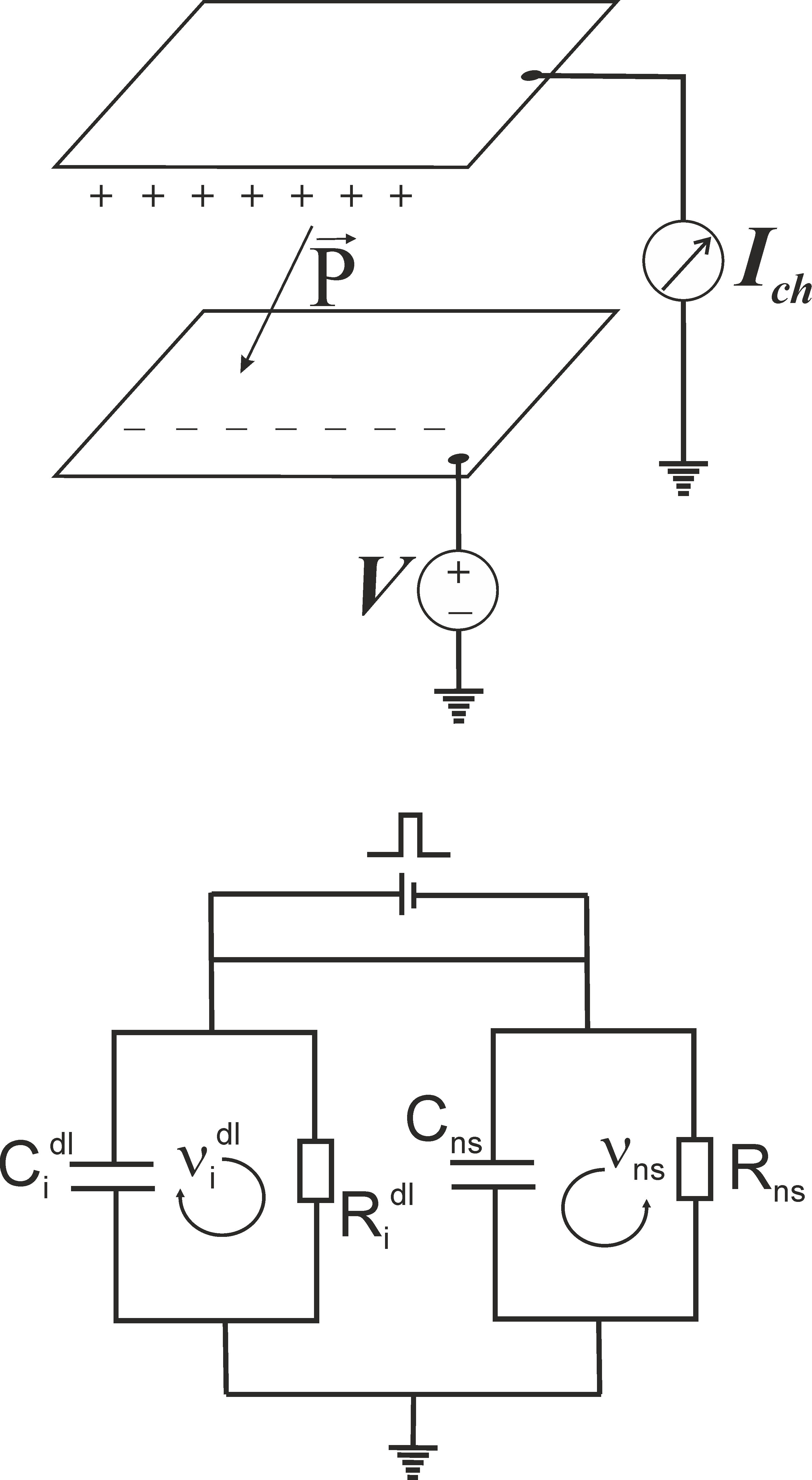}
\hspace{1cm}\includegraphics[width=8.cm,height=10.cm,angle=0]{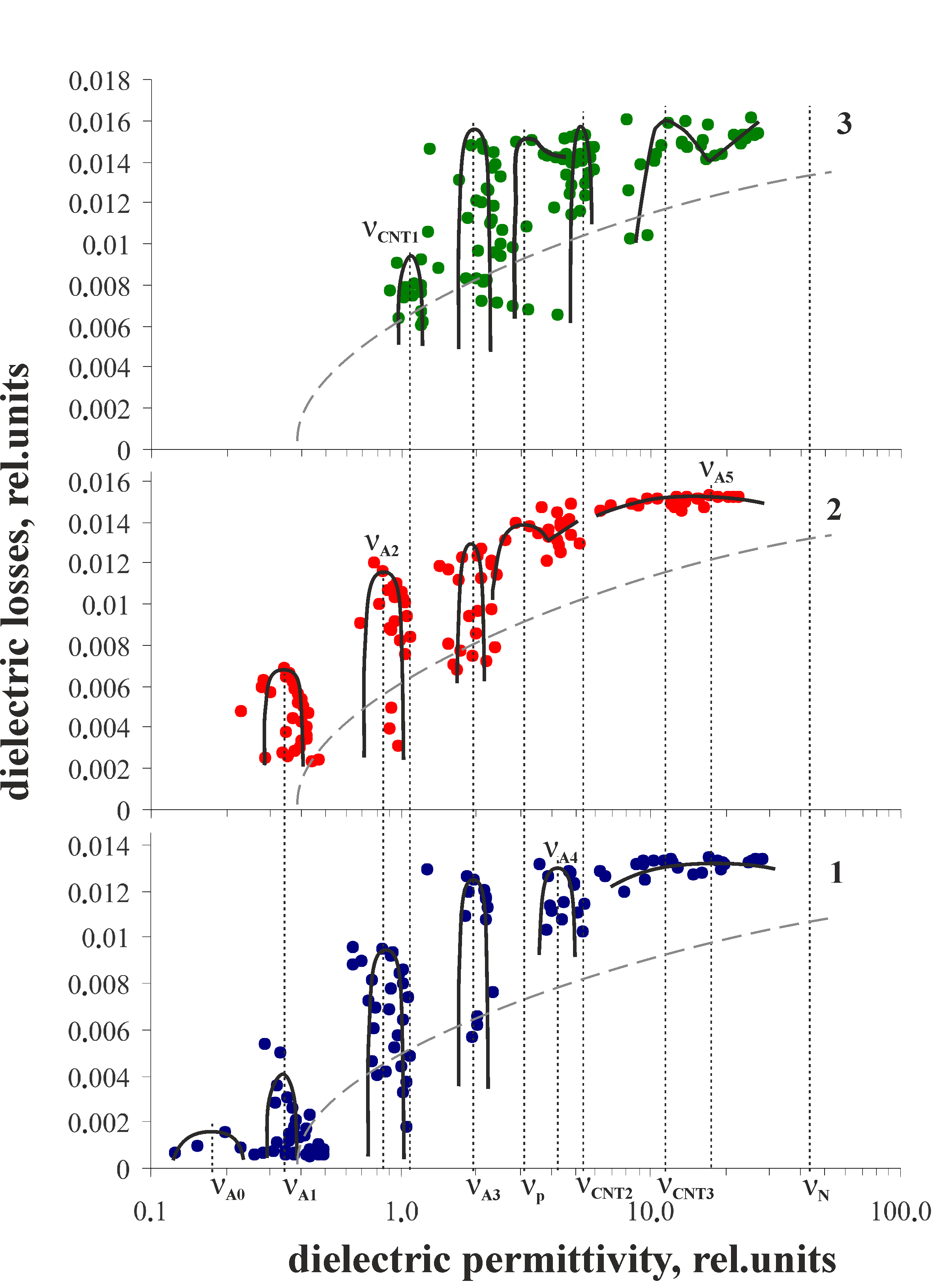}
\caption{(a) Cartoon of the scheme of the sensor operation based on a dielectric-polarization
relaxation process (up) and  an equivalent circuit diagram of the sensor (down). A square voltage
pulse with a bias is applied to the sensor to excite the autooscillations; $\vec P$ is the polarization
vector. (b) Dielectric spectra~``1'' for nanoporous AOA, ``2'', and ``3'' for nanoheterostructure based on
the 
nanocyclic compound Fe(II)DTP and on the decorated FWCNTs,
respectively. Quasiresonance frequencies of Cole-Cole plots in the
spectra are marked as $\nu _{Ai}$; $i = 0, \ldots, 5$ for AOA;
$\nu_P$ for the organometallic compound and $\nu_{\mbox{CNT}i}$,
$i = 1,2,3$ for FWCNTs. A Nyquist plot of impedance for the
measuring circuit is marked by a yellow dashed line with
$\nu_{\mbox{N}}$. The dielectric losses and permittivity were
measured  in inverse-capacity  units of pF$^{-1}$ and in power
units of V*V. } \label{opereation-principle}
\end{figure}

\section{Characterization of capacitive transducer and principle of its operation}

\subsection{Raman spectral study of DNA--CNT hybrid interactions}

A prominent CNT radial breathing  mode (RBM) of 298~cm$^{-1}$ is featured in Raman spectra  of the transducer
DNA--CNT hybrid layer (see Figs.~\ref{DNACNT}a and b). Let us estimate the CNTs diameters using the formula
\cite{Dresselhaus-Sato2010}
\begin{equation}
\omega_{RBM}={227\over d_t}\sqrt{1+C_{env}d_t^2} \label{environment-coef}
\end{equation}
where $C_{env}$ quantifies the environmental effect  on the RBM
frequency, $d_t$ is a  diameter of  single walled carbon nanotubes
(SWCNTs). We estimate approximately the $C_{env}$ at zero in the
case of the isolated original carboxylated few-walled CNTs. The
CNT diameter is estimated at  0.761~nm for the RBM of
298.28~cm$^{-1}$. CNTs with such diameter are single-walled ones.
Using the following formula 
\cite{Gangoli2014}: 
\begin{equation}
d_t={0.246\over\pi } \sqrt{\left(n^2+nm + m^2\right)}
\end{equation}
 one gets the index $(n,m)=(7,4)$ for the SWCNTs. 
It is  a sign of the fact 
that   they are of metal type because $ \mbox{mod}[(n-m),3] = 0$.


So-called   D and G  graphene Raman bands in the Raman spectra for the
broken-down (crumpled) LB-monolayer formed by the micellar
KRAS$_w$--FWCNTs hybrids and the micellar DNA$_{\mathrm{pl}}$--CNT
hybrids are very intensive  (see Fig.~\ref{DNACNT}).
The D peak is defect-activated one because it 
is highly intensive if the graphene structure is disturbed. 
The G peak is an in-plane vibrational mode that involves sp$^2$ hybridized carbon atoms
that comprises the graphene sheet. To find DNA  one utilizes a DNA intercalating dye, for example, propidium iodide (PI). PI
fluorescence grows when the PI molecules intercalate into the DNA
molecule.
It is used to indicate the presence of DNA.
As Fig.~\ref{DNACNT}c shows, such PI fluorescence is observed for
 the samples under investigation.

The crystal structure of the KRASw--FWCNT LB-monolayer or
high-ordered self-organized micellar structure change during
crumpling of the LB-monolayer or due to misfolding of the CNTs in
the inverse micelles. The high intensity of the D peaks of the
Raman spectra shown in Fig.~\ref{DNACNT} steams from the fact that
there are  isolated structural defects in the DNA--CNT hybrids.
However, since the metallic single walled CNTs are flexible they
can be arranged with a high degree of order. In this case, the D
and G peaks can be very weak in the Raman spectra of DNA--CNT
hybrids 
%
(see Figs.~\ref{fig1}(a, red curve) and (d, red and green curves 3
and 4)). Prominent peaks in these Raman spectra in the
Figs.~\ref{fig1}(a, red curve) and (d, red and green curves 3 and
4)) indicate the DNA molecular group vibrations   because their
frequencies coincide with all characteristic bands of micellar
dsDNAs in recorded Raman DNA spectra presented in
Figs.~\ref{fig1}(a, blue curve), (b)--(d) \cite{Babenko2020}.

Comparing the DNA Raman spectra shown in Fig.~\ref{fig1}a  before
and after their complexification with the high-ordered arrays of
the metallic CNTs one concludes that the intensity of the DNA
Raman bands is increased. This occurs in a result of the
complexification with the metallic CNTs and, correspondingly, the
CNTs can enhance the light scattering in the DNA layers. It
signifies that there are graphene patches which host Klein
resonances provided that the bulk graphene is doped. The Klein
resonance is excited by an electric field (for example, the
electric field of an electromagnetic pulse). Such areas are called
electrostatically-confined graphene $p-n$ ($n-p$) junctions
\cite{My2021PhysRevB}.
The confinement is explained by the occurrence of Klein resonances
as a result of Klein oblique tunneling of the massless graphene
charge carriers (holes or electrons) through these graphene
patches. It was proved in \cite{My2022JNPCSKleinRes} that the Klein resonances
have a topologically non-trivial origin. They exist due to
non-Abelian statistics of the graphene fermions, but not due to
the chirality of pseudo Dirac fermions. The obliquity of the
Klein-tunneling graphene charge carriers moving in the CNT samples
emerges as a result of the electrical dipole moment action from
the side of the DNA molecular groups.

%
%

The topologically-nontrivial graphene charge carriers are vortex
and antivortex Majorana excitations (modes) or their cores
in the graphene electron density 
\cite{My2022QuantumReport,myNPCS18-2015,NPCS18-2015GrushevskayaKrylovGaisyonokSerow,Taylor2016,our-symmetry2020}.
The Klein resonances and, correspondingly, the pairs of
Majorana modes are arranged on a crystal lattice if the
DNA-CNT hybrids form a superlattice structure. A number of the
Majorana modes is giant owing to the fact that a
topological-charge conservation law impedes their annihilation.
When the free pseupo-Majorana modes oscillate at the
eigenfrequencies of molecular DNA groups, plasmon resonances
arise. Their intensities are much higher than ones of the proper
DNA vibrations because the huge number of the Majorana
modes. It explains the CNT-enhanced scattering of light in the
DNAs which Raman spectra are presented
%
in Fig.~\ref{fig1}a.

The low intensity of the DNA Raman peaks in the broken (crumpled)
LB-monolayer is due to the fact that the isolated hybrid
structural defects play the role of disorderly-arranged impurities
residing in the Klein-resonance superlattice. To satisfy a Pauli
exclusion principle, 
the graphene pseudo-Majorana fermion pairs occupy higher
electronic impurity levels and, correspondingly,  prevent
electrons of the aromatic molecular groups  from passing to these
high-energy levels. This phenomena is called a fluorescence
quenching in the graphene plane. In this case the number of
negative (positive) charge carriers prevails over the number of
positive (negative) charge carriers in graphene. As a result, the
Dirac $K(K')$ point of the graphene Brillouin zone turns out to be
electrically charged. Oscillations of  C nuclei near the charged
Dirac $K(K')$ point  originate the D Raman peak because the
charged $K(K')$ point  plays the role of an electronic term. If
the Raman peak intensity is high enough, for example, as in the
Raman spectra shown in Fig.~\ref{DNACNT}, then weakly
intensive vibrations of the DNA molecular groups are barely
visible on its background.


DNA Raman spectra are strongly affected by deprotonation.
 Due to the latter   Raman spectral bands of DNA are widened even  for a dry
 DNA sample (see the Raman spectrum of the dry dsDNA$_{\mathrm{pl}}$ deposited on pure Si support in Fig.~\ref{fig1}e).
Hampering the deprotonation the stearic-acid shell of the inverse micelles
drastically narrows Raman spectral bands in the Raman spectra for the micellar dsDNA and dsDNA--CNT-hybrid
deposited on pure Si (make comparison between the Raman spectra presented in Figs.~\ref{fig1}a,b, and e).

Raman spectra for the micellar  dsDNA$_{\mathrm{pl}}$  dropped on
a Si surface hydrophilized by the H-DTP molecules are
significantly widened in comparison with the spectra of the
micellar DNAs deposited on the non-modified  Si surface  (compare
the two Raman spectra ``1'' and ``2'' shown in Fig.~\ref{fig1}d
with the Raman spectra in Figs.~\ref{fig1}(a, blue curve) and
(b)). However, the complexification of the dsDNA$_{\mathrm{pl}}$ with the
carbon nanotubes leads to the confinement of the placental DNA on
the hydrophobic CNT surface by the $\pi-\pi$  stacking
interactions and, correspondingly, to a DNA protonation. The
narrowing of the Raman spectrum of the micellar
DNA$_{\mathrm{pl}}$-CNT hybrids testifies the DNA protonation
(make comparisons between the values of spectral widths in the
Raman  spectra~``1''--``4'' shown in Fig.~\ref{fig1}d).


\subsection{Principle of operation of capacitive transducer   }

Electrophysical properties of the ultra-thin LB-films were studied
by means of the impedance spectroscopy methods as a variation in
dielectric polarization of a  near-electrode Helmholtz double
electrically charged layer formed at the interface between a
surface of the fabricated electrochemical sensor and water.
The non-faradaic impedimetric sensor operates on surface-polarization-decreasing effects
which are originated by the conducting ultra-thin LB-films when shielding the  near-electrode Helmholtz layer. A scheme
of the sensor operation is presented in Fig.~\ref{opereation-principle}(a). As
Fig.~\ref{opereation-principle}(a,~up) shows, the discharging
(charging) current $I_{ch}$ produces the bias $V$ in the RC oscillator.
A polarization $P$ of near-electrode layers is proportional to the applied bias $V$:
$\Pi P\propto V$, where  $\Pi$  is a polarizability.
On the other hand,  $P$ is the induced charge density 
$\delta n$ due to the charging (discharging) of the capacitor so
that the charging (discharging) current 
$I_{ch}$ is proportional to $\delta n$: $P\sim \delta n \sim I_{ch}$.
Since $P$ is determined through a complex dielectric permittivity $\epsilon$  by the following
formula: $\epsilon =1 -\Pi$, one gets that
\begin{equation}
I_{ch} \propto {V\over 1- \epsilon}. \label{charging-current}
\end{equation}
The dielectric losses $\epsilon ''$ due to the dielectric-polarization relaxation
process occur in a frequency range where the sensor capacitance $C$ falls down.
The $\epsilon ''$ is the imaginary part of the complex dielectric permittivity:
$\epsilon = \epsilon' + i\epsilon '' $, and, correspondingly, presents a resistivity because
\cite{Kraft-Ropke}
\begin{equation}
\epsilon  =1+i {\sigma\over \epsilon_0 \omega} .
\end{equation}
Here $\sigma$ is the conductivity, $\epsilon '$ is the dielectric permittivity,
 $\omega $ is the cyclic frequency, $\epsilon_0$ is the electric constant. It follows from the last equation
and Eq.~\ref{charging-current} that
\begin{equation}
I_{ch}\sim \Im m\ {V\over 1 - \epsilon } \sim  \Re e {\epsilon_0 \omega V\over \sigma }
\sim {V\over \epsilon '' }. \label{charging-current1}
\end{equation}
But, an energy $CV^2 / 2$ stored at (released from) the
capacitance $C$ is determined by the following expression: $ CV^2
/ 2 \sim  I_{ch} V$. Substituting the expression
(\ref{charging-current1}) for the $I_{ch}$ into the expression for
the energy one can assume that $\epsilon ''\sim 1/C$. The signal
power $W$ recorded by the Fourier analyzer is proportional to the
real part of the complex dielectric permittivity $\epsilon ' $
because $W$ is the stored energy determined by the following
equation: $W\sim \left| \vec D \times \vec H \right| $ with the
electric displacement vector $\vec D = \epsilon ' E $, where $\vec
E$ and $\vec H$ are electric and magnetic fields vectors,
respectively.

Let us suppose that $N$ dipole-polarization processes proceed inside the Helmholtz layer
emerging near the $N$ electrode pairs. It signifies that an equivalent circuit diagram of both the
capacitor-charging (capacitor-discharging) and polarization processes consists of $(N+1)$ RC
circuits. The equivalent circuit of the sensor presents in Fig.~\ref{opereation-principle}(a,~down).
As one can see forced oscillations with a frequency $\nu_i$ can be generated in
the $i$-th parallel $R_i^{dl}C_i^{dl}$ circuit for which the double-layer resistance $R_i^{dl}\equiv R_i$,
$i=1,\ldots, N$ is much more
than the instrument resistance $R_{ns}$ in some frequency range $\nu_i\pm \Delta _i$. If $R_i \ll R_{ns} $
then the forced oscillations with a frequency $\nu_{ns}$ are generated in the properly
measuring  $R_{ns}C_{ns}$ circuit. Thus, the dependence of $1/C$ on $W$ will present $N$ Cole--Cole
plots and one Nyquist impedance plot for parallel connection of the capacitor to the
instrument resistor $R_{ns}$ (see Fig.~\ref{opereation-principle}b).
Since $R_i \gg R_{ns} $ the capacitance $C$ of the sensor can be calculated by the formula $C = 1/(2\pi R_{ns} f)$.

The charge density of the graphene monolayers deposited on the
interdigital electrode structure in deionized water can be
indirectly resonantly swung by the low-frequency resonant decay of
the hydrated ion complexes, which are in an excited state, and
subsequent association of the ions OH$^-$ and H$^+$ into water
molecules. Meanwhile,  emitting  electromagnetic quanta the water
molecules pass to a low-lying state. The electromagnetic field
excites the graphene-charge-density oscillations. Resonating with
dipole moments of the hydrated complexes of the near electrode
double charged layer, the graphene plasmon oscillations decay the
hydrated complexes in the Helmholtz double layer and,
correspondingly, screen effectively the electrodes \cite{my2017AdvMatLet}.

Dielectric spectra of the AOA isolating  layer, Fe(II)DTP LB-film,
and FWCNT LB-arrays under investigation are shown in
Fig.~\ref{opereation-principle}b. The dielectric spectra of AOA
do not have specific dielectric Cole-Cole plots with a Warburg
diffusion element or Warburg impedance, $Z_W$. Since the Warburg
impedance is absent for AOA and, consequently,
aluminium-electrodes oxidation does not happen, the fabricated
nanoheterostructures are stable.

The $ Z_W$ impedance element is created by the diffusion layer in
which the electrochemical reactions with a mass transfer proceed due to electrochemical reactions.
The impedance measurements for the metal-containing dithinylpyrrole LB-films
entering the Helmholtz double layer indicate
the depressed semi-circle with the Warburg diffusion tail in the Cole-Cole plot
of the dielectric spectrum ``2'' shown in Fig.~\ref{opereation-principle}b.
For the dithienylpyrrole LB-films, this oxidation-reduction
potential (potential stipulated by a change of molecular-group
charge state) arises due to self-redox activity of pyrrole residue
leading to the transport of electrical charge along the chain of conjugated double bonds.
The FWCNT LB-arrays  
are redox active themselves also because a Warburg impedance element
is visible in the corresponding Cole-Cole plot from the dielectric spectrum ``3'' shown in
Fig.~\ref{opereation-principle}b. The change of charge state of
valleys $K(K')$ (Dirac points) of the graphene Brillouin zone in
the momentum space is associated with the mass transfer that
Kitaev-like chains with Majorana end states emerge. Meanwhile,
plasma oscillations of produced electron-hole pairs shield
electric field of charged electrodes.

In \cite{3}  a method to detect the DNA hybridization  leading to a homoduplex formation at low temperatures
 has been proposed. This method operates on  the enhancement of the
electric-field FWCNT screening  when the FWCNTs are linked to each
other by dsDNA and as a result a high-conductivity network is
created.

\subsection{Dielectric spectral analysis of low-temperature DNA hybridization }
\begin{figure}[htbp]
\centering
(a)  \hspace{6.3cm} (b) \\
\includegraphics[width=7.cm,height=6.cm,angle=0]{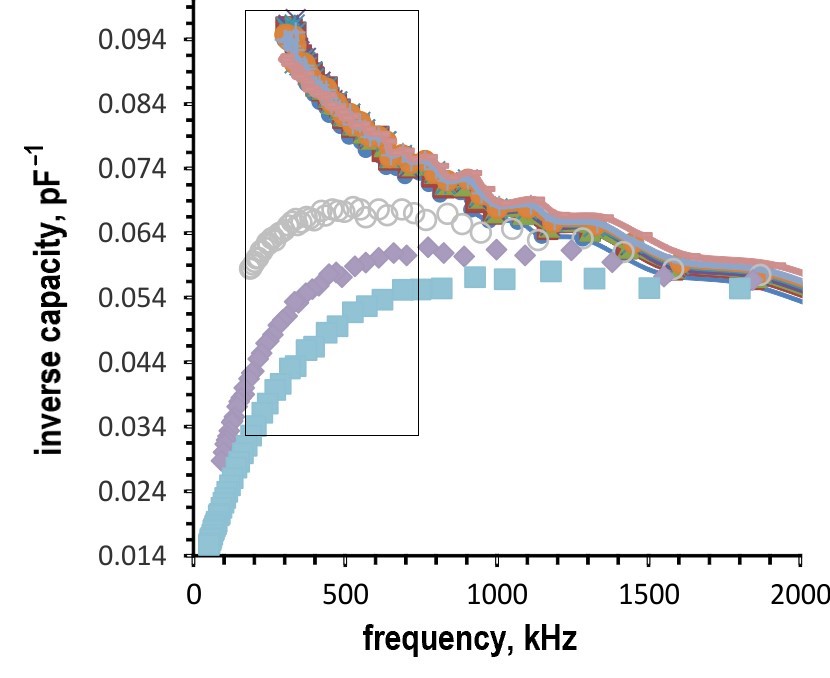}
\includegraphics[width=7.cm,height=6.cm,angle=0]{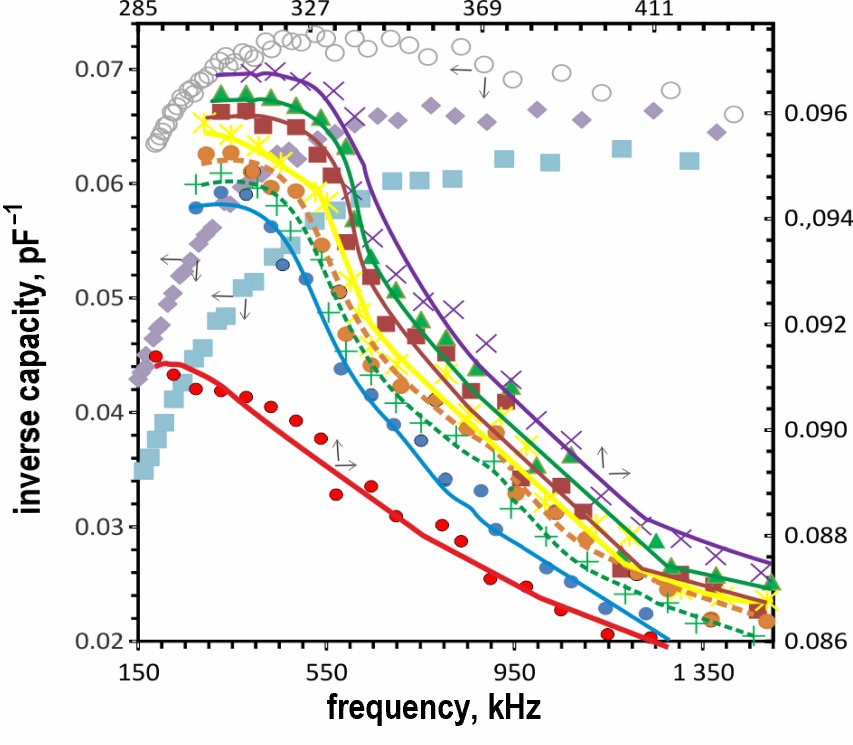}
\caption{Bode plots 
at different temperatures for P3/N3 homoduplexes. Brown, green, yellow, red and blue solid curves fit
the data labeled by symbols of corresponding color: brown squares
``$\blacksquare$'' for 30~$^\circ $C, green triangles
``$\blacktriangle$'' for 32~$^\circ $C, violet ``$\times$'' for
34~$^\circ $C, yellow ``$\ast$'' for 36~$^\circ$C, red circles
``$\bullet$'' for 50~$^\circ $C, blue circles ``$\bullet$'' for
RT. Light brown, green dashed curves fit the data labeled by
symbols of same color: light brown circles ``$\bullet$'' for
38~$^\circ $C, green ``$+$'' for 40~$^\circ $C. The data labeled
by open gray circles ``$\bigcirc$'', light violet rhombuses
``$\lozenge$'', and light blue squares ``$\square$'' present
dielectric loss at 60, 65, and 70~$^\circ $C, respectively. Inset
(b): the designed fragment of the figure (a)
is shown on a large scale. } \label{fig2}
\end{figure}

Now, let us study  a toehold exchange reaction between the dsDNA
``P3/W3''  and  the ssDNA ``N3'' proceeds in the DNA chip. The
formation of the P3/N3 homoduplex in the result of the toehold
exchange reaction between the dsDNA  ``P3/W3''  and  the ssDNA
``N3'' is detected as an rise of the dielectric losses, $1/C$.
When hybridizing  the toehold DNA probes the N3 ssDNA target in
the temperature range from RT to 34~$^\circ $C because the
dielectric-loss spectral peak grows in this temperature range due
to increasing the electrode shielding degree by the sensing
covering after the homoduplex formation as Fig.~\ref{fig2} shows.

The Bode peak intensities vary weakly for the temperatures of 30,
32, 34, 36, 38, 40, 50 $^\circ $C and largely enough for 60, 65,
70 $^\circ $C in comparison with the intensity of the peak at RT
(see Fig.~\ref{fig2}). One observes an abrupt change  (drop) of
the dielectric losses at 60~$^\circ $C. The drastic drop of the
spectral peak intensity  at 60~$^\circ $C indicates that defects
appear in the sensitive coating  due to a disruption of the
stacking $\pi - \pi$ interaction and of the hydrogen bonds between
complementary nucleotide pairs in the DNA homoduplexes. This
screening-effect attenuation testifies that dsDNA melting occurs.

A similar stepwise drop of the dielectric losses, but on a smaller
scale, is observed at 50~$^\circ $C because the intensity of the
Bode plot peak at 50~$^\circ $C drops significantly more than the
intensities of the nearby Bode plots for the following
temperatures: 36, 38, and 40~$^\circ $C.
%

Such universal scaling dielectric losses behavior allows us
suggest that  the systems undergoes a first-order phase
transition.
Thus the existence of the temperature scaling indicates that the
process of dsDNA dissociation on the surface proceeds similarly to
a phase transition of the first order.

The DNA-melting temperature $T_m$ is indicated as the temperature
at which  the effect of the electrode-electric-field screening
disappear. The dielectric losses increase with the increase in
temperature  up to the temperature of 34~$^\circ$C. Then, with the
increase in temperature to the temperature of 36~$^\circ$C the
intensity of the dielectric losses are reduced. It means that a
value of $T_m$ is close to 36~$^\circ$C.

\begin{figure}[htbp]
\centering
\includegraphics[width=13.cm,height=7.cm,angle=0]{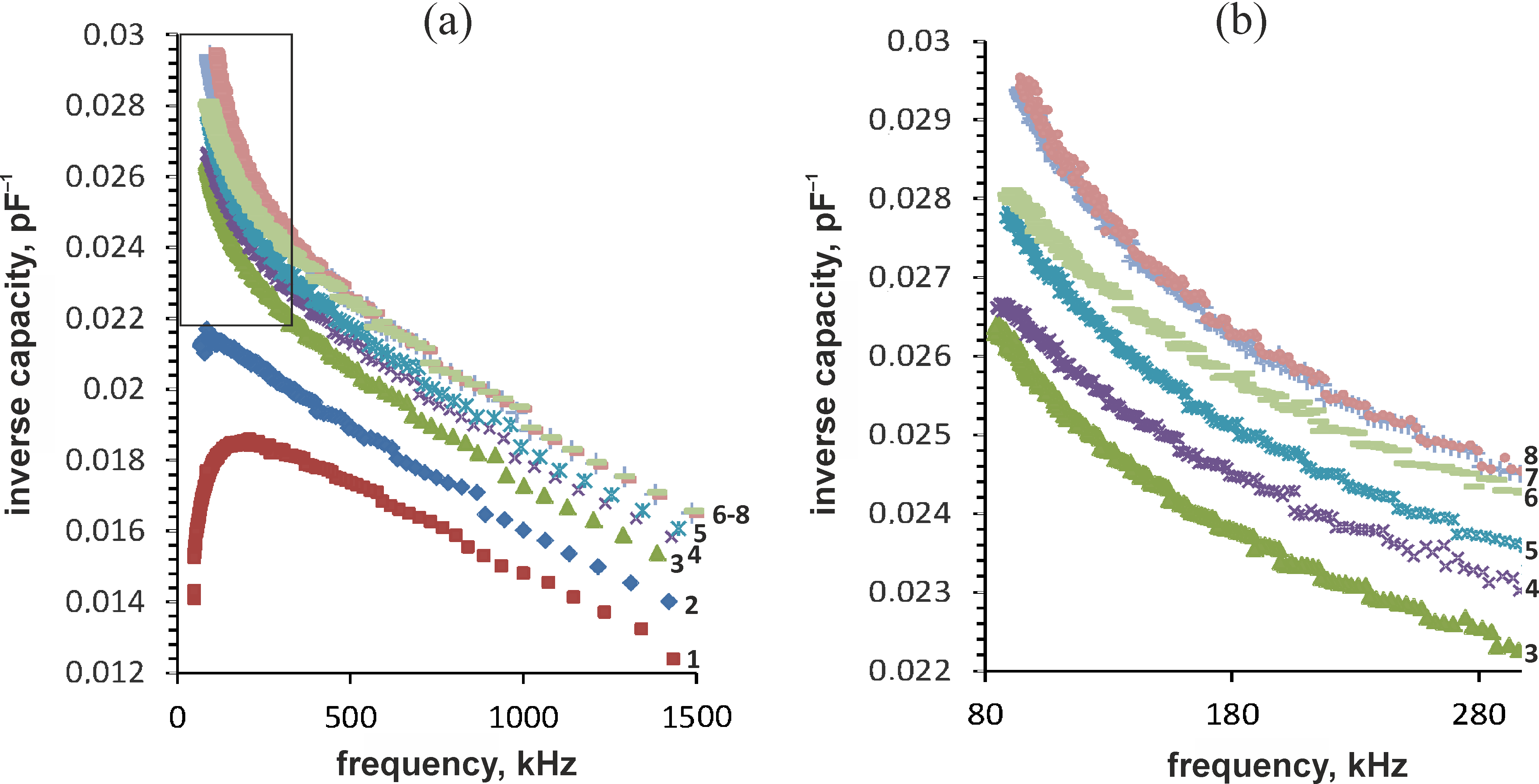}
\caption{Bode plots 
at different concentration of the target DNA
after hybridization between the N3 or M3 ssDNA target and toehold probe P3/W3 : the plot ``1'' (brown squares ``$\blacksquare$'')
for the pure sensor; the plot ``2'' (blue rhombus
``$\blacklozenge$'') for the DTP sensor coating; the plot ``3''
(green triangles ``$\blacktriangle$'') for the polyG-FWCNT
hybrids; the plot ``4'' (violet criss-crosses ``$\times$'') for
the P3/W3 probe; the plots ``5''--``7'' for  the P3/N3
homoduplexes forming  at the target concentration of $10^{-18}$, 
$10^{-14}$, 
and $10^{-12}$ mol/$\mu$L,
respectively; the plot ``8''  for the mutant M3 ssDNA target
hybridized with the P3/W3 toehold  at the M3 concentration of
$10^{-18}$ mol/$\mu$L. Inset (b): the designed fragment of the
figure (a) is shown on a large scale. The hybridization proceed at
RT.} \label{fig4}
\end{figure}

\begin{figure}[htbp]
\centering
(a) \hspace{7.5cm} (b)\\
\includegraphics[width=7.cm,height=7.cm,angle=0]{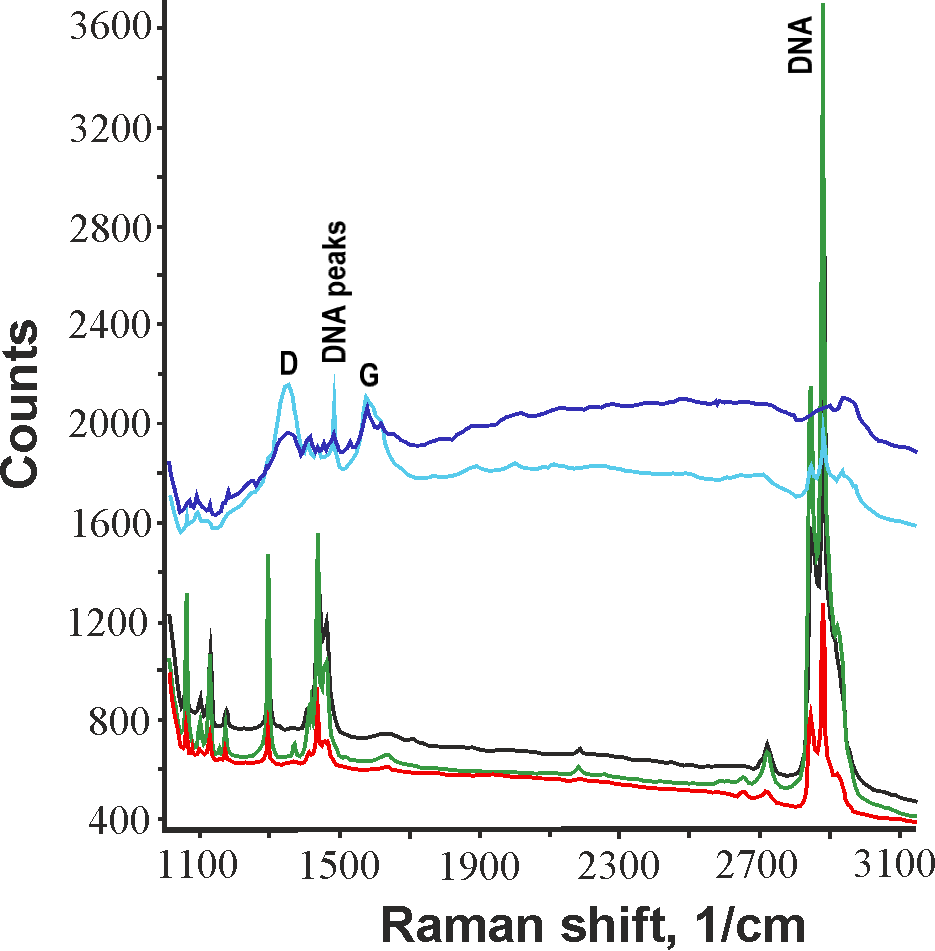} \hspace{0.8cm}
\includegraphics[width=7.5cm, height=7.0cm, angle=0]{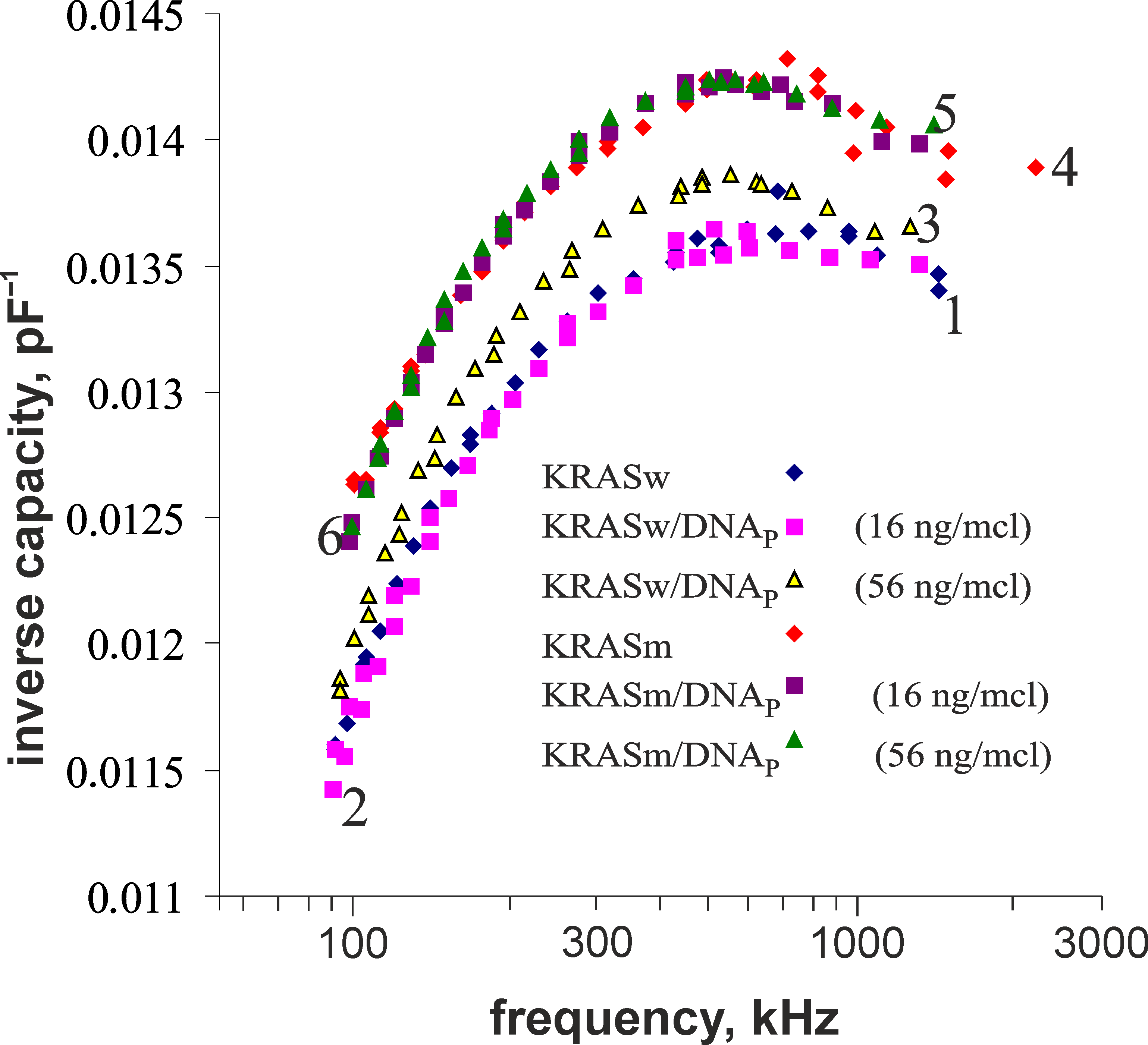}\\
(c) \hspace{7.5cm} (d)\\
\includegraphics[width=7.5cm, height=7.0cm, angle=0]{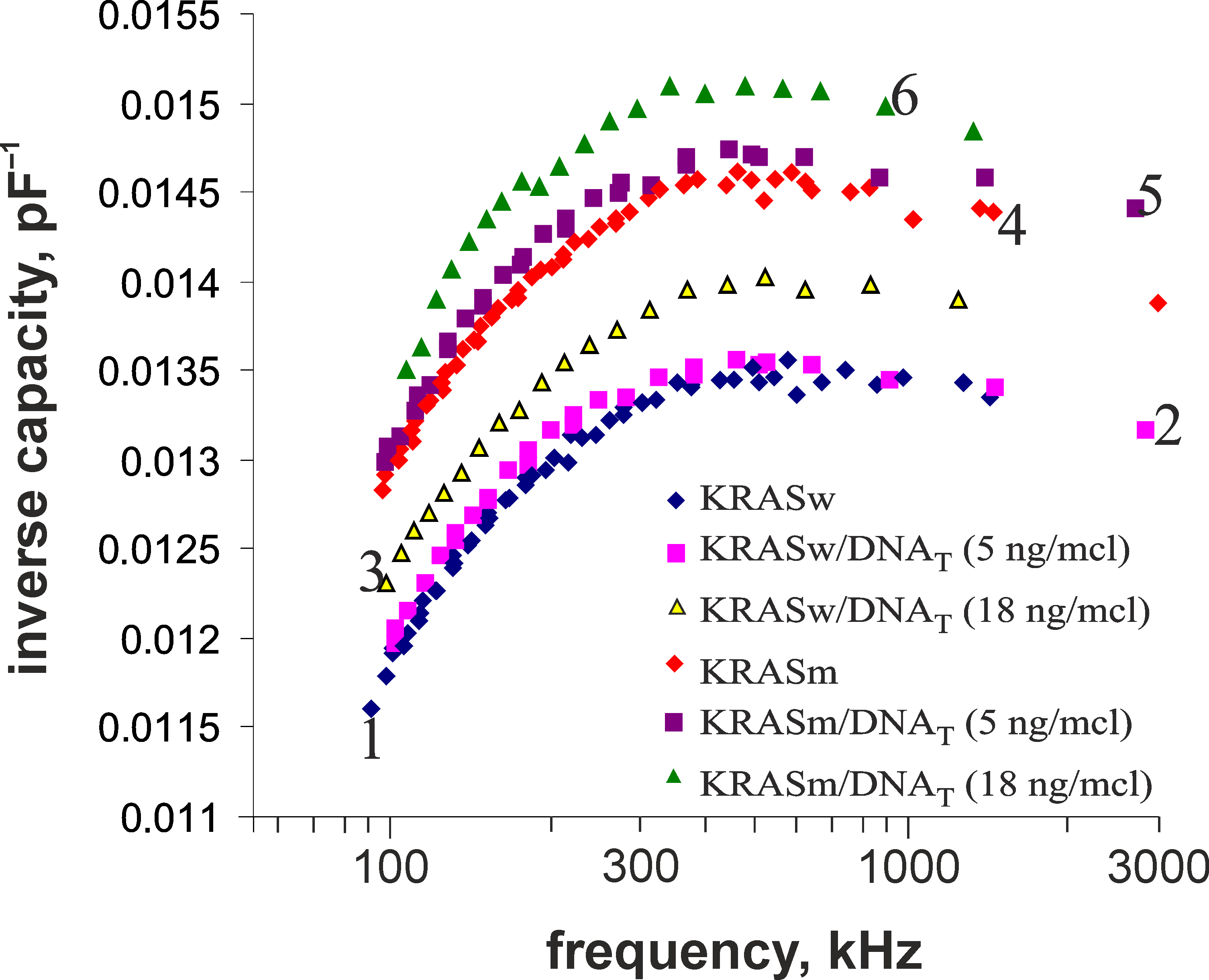} \hspace{0.8cm}
\includegraphics[width=8.cm, height=9.0cm, angle=0]{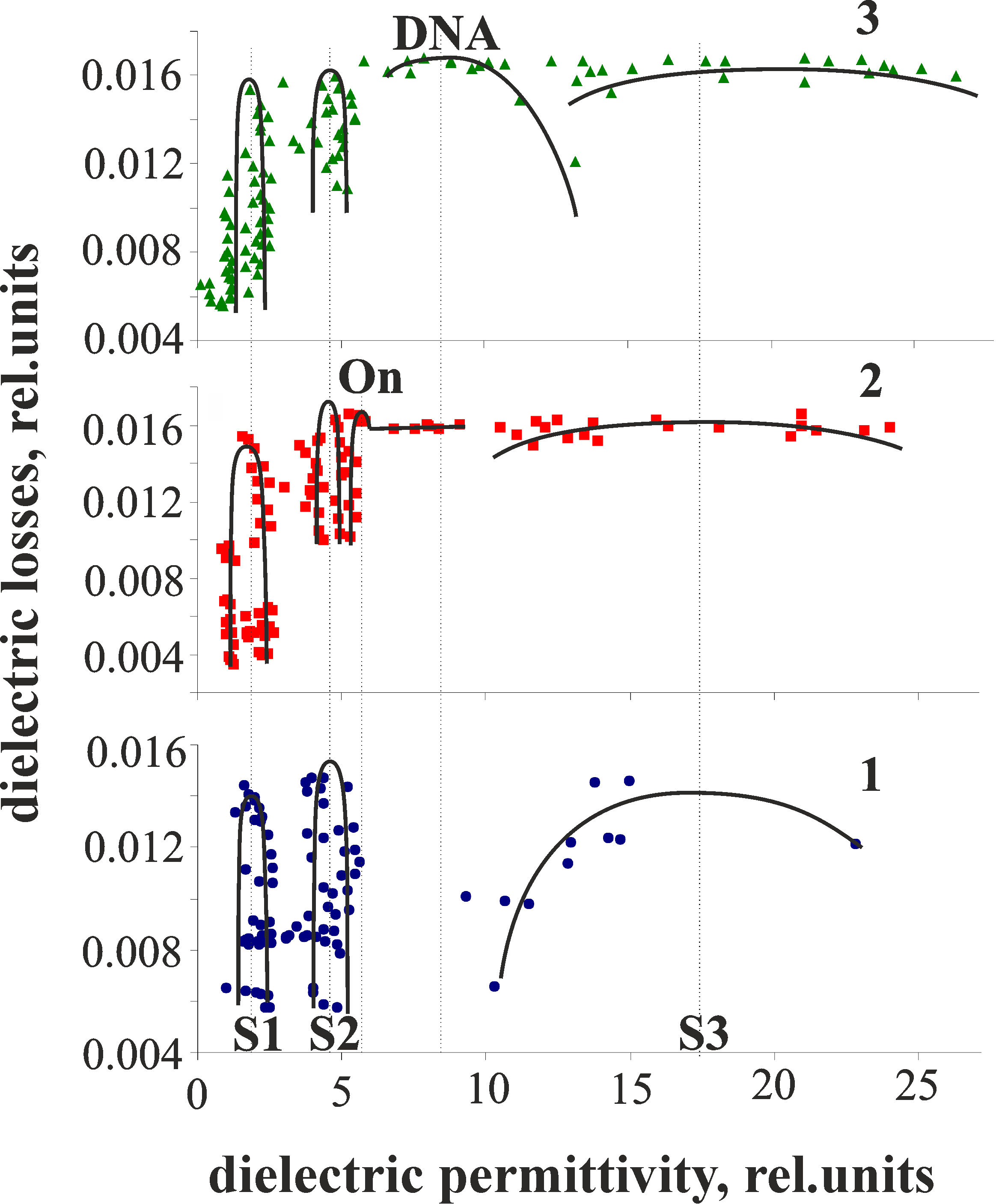}
\caption{Raman spectral (a) and EIS (b--d) genotypings. (a) Raman
spectra of inverse  micelles with DNA$_{\mathrm{C6}}$ isolated
from C6-line cells (black curve), of imperfect micellar FWCNTs
covered by DNA molecules hybridized with KRAS$_w$- or
KRAS$_m$-oligonucleotides (light- and dark blue curves,
respectively), and of high-ordered micellar FWCNTs covered by DNA
molecules hybridized with KRAS$_w$- or KRAS$_m$-oligonucleotides
(green and red  curves, respectively). The spectra were recorded
at laser excitation wavelengths of 532~nm; the laser power and
collected time were used for the specimen excitation: 2~mW and
60~s, respectively. (b--c) Bode plots for sensors before (1, 4)
and after (2,3,5,6) hybridization between perfectly matched and
single-nucleotide mismatched ssDNA probe with native
DNA$_{\mathrm{pl}}$ (b) and DNA$_{\mathrm{CRC}}$ (c). DNA
concentrations used are indicated in brackets.
(d) 
Dielectric spectra:  ``1'' and ``2''  for the DNA-sensor  with 
$KRAS_w$ probe and  $KRAS_w$--ssDNA$_{lp}$/ssDNA$_{Wm}$ 
toehold probe system,  respectively; ``3'' after binding  with the target ssDNA$_{lm}$.
Cole-Cole plots are labeled by S$i,\ i=1, 2, 3$ for the DNA-sensor with $KRAS_w$ only; ``On'' for  the dsDNA toehold probe,
and ``DNA'' for the homoduplexes, respectively.
The dielectric losses and permittivity were measured  in
inverse-capacity  units of pF$^{-1}$ and in power   units of V*V.
} \label{fig3-inverCap-dielectrSpectr}
\end{figure}

{\underline{Sensitivity of label-free electrochemical impedimetric
DNA-chip}.}

As one can see in Fig.~\ref{fig4}, the binding of the DNA probe with
hydrophobic sensor surface occurs as the capacity decrease is
observed. This screening effect is provided by $\pi$--$\pi$
interaction between the DNA probe nucleotide sequence and the surface
of the carbon nanotubes. The capacity decrease is of, for example, 2~pF at 150~kHz.
Accordingly to the experimental data presented in Fig.~\ref{fig4}, the replacement of the ``weak'' inosine-contained chain
``W3'' entering the toehold probe ``P3/W3'' by the perfect-matched target ssDNA ``N3'' occurs at room temperature.
The screening effect as a result of the complementary hybridization holds in the all frequency range.
The sensor electrodes are shielded  both at picomolar and
attomolar concentrations of the DNA target.

When adding the single-mismatched target ssDNA ``M3''  increases
the sensor capacity $C$ significantly that is stipulated by the
bulk charge density increase. The large bulk charge density
appears because the non-specific links between the
non-complementary target and probe DNA molecules and, as a result,
gapping between the DNA probe and target that leads to dipole
polarization and ionization of the sugar phosphate backbone.

After washing of the P3/M3 heteroduplexes, the capacity of the
sensor returns to its pre-hybridization value. It signifies that
the heteroduplexes do not penetrate into the nanopores and are
held on the sensor surface by electrostatic forces if there is no
first-order phase transition. Correspondingly, the heteroduplexes
are easily washed away because of the weakness of the
electrostatic forces.


So, a network is created by linking together  FWCNTs through  dsDNA homoduplexes that results in the increase of
sensor-shielding degree.


\section{KRAS-gene sequencing}

\subsection{Micellar genotyping based on quenching effect}

Forming a high-conducting network of CNTs linked by DNA molecules
the complexification of highly-ordered CNT arrays with DNA does
not impair but vice versa improves the ability of the CNT arrays
to screen electromagnetic fields. Meanwhile, the ssDNA--CNT
hybrids have to be disordered enough so that when occupying the
impurity energy levels the graphene pseudo-Majorana  charge
carriers would quench the light scattering in DNA.

A perfect network of  DNA--CNT hybrids  
is formed by links between the KRAS$_m$-dsDNA$_{\mathrm{C6}}$--CNT hybrids
but not by links  between the KRAS$_w$-dsDNA$_{\mathrm{C6}}$--CNT hybrids
because the intensity of the Raman D-peak is significantly less 
for the KRAS$_m$-dsDNA$_{\mathrm{C6}}$ duplexes (compare Raman spectra in
light- and dark-blue lines in Fig.~\ref{fig3-inverCap-dielectrSpectr}a). The attenuation 
of the quenching effect indicates that the
KRAS$_m$-dsDNA$_{\mathrm{C6}}$--CNT network is more perfect and when
hybridizing the  KRAS$_m$ primers and dsDNA$_{\mathrm{C6}}$ molecules form
homoduplexes. Moreover, the intensity 
of the KRAS$_w$-dsDNA$_{\mathrm{C6}}$-heteroduplex Raman bands (see  the
frequency region from 2700 to  3100~cm$^{-1}$ in the Raman spectra
in light-blue lines in Fig.~\ref{fig3-inverCap-dielectrSpectr}a) is significantly higher
than one of the KRAS$_m$-dsDNA$_{\mathrm{C6}}$-homoduplex Raman bands (see
the frequency region from 2700 to  3100~cm$^{-1}$ in the Raman
spectra in dark-blue lines in Fig.~\ref{fig3-inverCap-dielectrSpectr}a). It testifies that
DNA interior of the micelles hosting KRAS$_m$-dsDNA$_{\mathrm{C6}}$
duplexes is shaded from the laser radiation by  the
KRAS$_m$-dsDNA$_{\mathrm{C6}}$--CNT network better than the
KRAS$_w$-dsDNA$_{\mathrm{C6}}$--CNT network due to the fact that surface
of the KRAS$_m$-dsDNA$_{\mathrm{C6}}$--CNT micelles is covered by 
the highly screening network created in the result of the
complementary hybridization.

Thus, creating  the network the homoduplexes  link the FWCNT
together so that the  degree of ordering 
of the micellar DNA-CNT hybrids increases after  hybridization.

\subsection{Micellar genotyping on plasmon-resonance effect}

The micellar genotyping on the plasmon-resonance effect is
performed in a following way. One detects a light scattering
enhancement in duplexes between the target and probe DNAs and
subsequently a dependence of this enhancement on the DNA--CNT
hybrid type  is analyzed.

 The plasmon resonance as well as the screening effect occurs at
high enough density of the charge carriers in the graphene valleys
$K,K'$. As the non-specific linked 
target DNA is kept on a long enough distance from the micelles
surface, the light scattering 
in the non-complementary hybridized target DNA is not quenched
because no transitions of
the graphene charge carriers to non-occupied impurity energy levels occur. 
The higher intensity of the DNA Raman bands for the KRAS$_w$
probe--target DNA duplexes hybridized with FWCNTs in comparison
with the intensity of the Raman spectrum of the micelles with the
original DNA testifies the FWCNT-enhanced  light scattering for
the DNA (compare black and green curves in Fig.~\ref{fig3-inverCap-dielectrSpectr}a). To
produce the plasmon oscillation DNA
molecules should reside outside the 
micelles surface so that the DNA-molecular groups vibrate out of
graphene plane and, accordingly, are not quenched.

The intensity of the micellar DNA$_{\mathrm{C6}}$--FWCNT-hybrid Raman bands
for the  homoduplexes, dsDNA$_{\mathrm{C6}}$-KRAS$_m$, of the target
dsDNA$_{\mathrm{C6}}$ with the mutant-type probe, KRAS$_m$,  are less than
for the heteroduplexes, dsDNA$_{\mathrm{C6}}$-KRAS$_w$, of the target
dsDNA$_{\mathrm{C6}}$ with the single-base mismatched oligonucleotide (as
one can see from comparison between the Raman spectra in red and
green lines in Fig.~\ref{fig3-inverCap-dielectrSpectr}a). It signifies that a perfect
high-conducting network of the homoduplex--CNT hybrids is produced
and when arranging on the micelles surface a  patch of the network
is shaded KRAS$_m$-dsDNA$_{\mathrm{CRC}}$--CNT hybrids residing in the
inside of the micelles   from the laser radiation.

\subsection{Electrochemical genotyping based on screening effects of crystalline CNT LB-arrays}

We present a micellar genotyping electrochemical technology to
discriminate a single nucleotide mutation in the KRAS gene (2 exon,
12 codon, G/A). To recognize the mutant and wild alleles, it was
necessary to use two sensors. The wild-type allele detection probe
was placed on the surface of the first sensor. The probe for
detection of the mutant type allele was located on the surface of
the second sensor.

This technology is based on the screening effect leading to that the capacity values for a wide frequency
range decrease as a result of the complementary hybridization between
the target and the probe DNA sequences. The degree of sensor
electrodes shielding increases after complementary hybridization
between the wild-type dsDNA$_{\mathrm{pl}}$ and $KRAS_w$ probe. At that time
the capacity of sensor decreases accordingly to its Bode plots for
different dsDNA$_{\mathrm{pl}}$ concentrations (see Fig.~\ref{fig3-inverCap-dielectrSpectr}b). No binding
between the dsDNA$_{\mathrm{pl}}$ target and $ KRAS_m$ probe occurs. The transducer
do not response on the KRAS mutation as Fig.~\ref{fig3-inverCap-dielectrSpectr}b shows. The
shielding effect is observed for the dsDNA$_{\mathrm{CRC}}$ target hybridized
both with $ KRAS_m$ and $KRAS_w$ probes. The capacity decreases in both
cases (see Fig.~\ref{fig3-inverCap-dielectrSpectr}c). It signifies that both wild- and mutant-type of KRAS-oncogene were detected.

Thus, the sequencing method has allowed us to diagnose  allele SNP
of the KRAS-oncogene in the genome of the colorectal cancer
tissue.
It testifies allele SNP of KRAS-oncogene in the genome of the
colorectal cancer tissue.

To study the selectivity 
of the DNA nanosensor 
a DNA chip has been designed. Two probes,
the KRAS$_w$--CNT LB-hybrids and the 47-base oligonucleotide
toehold probes of mutant type presented in Table~\ref{tab1}, have
been located on the chip surface. A mass transfer for the 47-base
toehold probe--FWCNT hybrids is observed, as the dielectric
spectrum of the EIS-transducer including the dsDNA toehold
probe--FWCNT hybrids is characterized by the presence of a Warburg
impedance in the Cole--Cole plot labeled by ``On'' for the
dielectric spectrum ``2'' shown in
Fig.~\ref{fig3-inverCap-dielectrSpectr}d.  
It testifies redox activity of the dsDNA probe on the FWCNT surface.
Fig.~\ref{fig3-inverCap-dielectrSpectr}(d) shows also that a Cole--Cole plot  of
homoduplex (Cole-Cole plot labeled by ``DNA'' in the dielectric
spectrum ``3'') appears after specific complementary
hybridization. Any signs of transducer response on the 
heteroduplex formation are not observed. 

Thus, DNA molecules being non-specific bonded 
with DNA probes are easily washed and do not contribute in the
transducer response. 

\section{Discussion and conclusion}

So, the experimental studies performed have revealed 
that no hybrid formation can  occur due to 
the stacking interactions between $\pi($p$_z)$ electrons of carbon
nanotubes and DNA molecules only. The observed 
CNT-enhanced scattering of light in dsDNA and shielding effects 
of DNA hybridization are explained by the topologically non-trivial Majorana nature of the graphene charge carriers.
the graphene topologically non-trivial charge carriers they transit to DNA non-occupied energy levels.
The stacking $\pi-\pi$ interaction can be originated by the Klein
resonances which are created  by Klein-tunneling graphene
topologically non-trivial charge carriers and subsequently transit
to DNA non-occupied energy levels.

 The conservation law of
topological charge forbids annihilation between the
pseudo-Majorana graphene charge carriers and the DNA charge
carriers. It signifies that a lifetime of the pseudo-Majorana
particles on the overlying impurity DNA levels is long. The
graphene charge carriers excited to the impurity energy
levels do not annihilate 
with 
electrons  or holes of the impurity because their statistics is
non-Abelian. The disappearance (quenching) of the light scattering
stems from the non-Abelian statistics owing to impossibility to
excite DNA electrons on the overlying levels due to the Pauli
exclusion principle.

Sharp decrease of the intensity 
of the CNT Raman bands up to it fully disappearing 
occurs in strong electrical fields of superlattices formed by the
DNA--CNT hybrids. Meanwhile, the scattering of light in the
aromatic molecular DNA groups residing away from the graphene
plane is enhanced because the light is scattered on intensive
graphene-charge-density oscillations emerging in resonance with
low-intensity vibrations of the aromatic molecular groups. The conducting properties 
of the DNA transducer are improved at the 
complementary hybridization. The screening of the light scattering
by the homoduplex--CNT hybrids is so amazingly effective that the
incident electromagnetic quanta do not reach and, accordingly, do
not excite the DNA vibration with subsequent generation of the CNT
plasma oscillations at eigenfrequencies  of DNA vibrations.

The DNA--CNT hybrid high-ordered coverings effectively shield 
near-electrode double charged layers. The new high-performance DNA
nanosensor which operation is based on a difference in efficiency
of the shielding 
by the ssDNA--CNT hybrid networks before hybridization and
dsDNA--CNT networks forming after complementary hybridization have
 been developed to genotype DNA sequences from fragmented FFPE nucleic acid samples with limit detection up
to attomolar concentrations.

Finalizing, dielectric and Raman spectral studies of DNA
hybridization on surface of metallic single-walled carbon
nanotubes decorated by organometallic compound and  suspended on
nanopores have been performed. It has been shown that self-redox
activity of the carboxylated CNTs decorated by the nanocyclic
orga\-no\-me\-tal\-lic complexes can be used  to discriminate
single-nucleotide poly\-mor\-phism of colorectal tumor genome.

We offer the two DNA-sequencing methods based on shielding effects
or enhancement of light scattering by DNA--single-walled metallic
carbon nanotube hybrids. Both the high-performance genotyping
assays operate at room temperature. The  impedimetric DNA sensors
of non-faradaic type based on the screening effect can be more
sensitive than the Raman optical transducer based on the quenching
effect due to liability of the Raman transducer  parameters to
environmental influence.
 Our high performance DNA sensor will be at an advantage over the plasmon-resonance sequencer
when performing single-molecule allele discrimination of the
genome of metastatic tumor cells at early and first stages of
cancer.



\end{document}